%% file: eprint-2.tex
\documentclass[12pt]{article}
\usepackage{graphicx}

\newcommand\pubnumber{NuPhys2016-Turner}
\newcommand\pubdate{\today}

\def\sussex{$^1$Department of Physics and Astronomy\\
University of Sussex, Brighton, UK}

\def\oxford{$^2$Department of Physics\\
University of Oxford, Oxford, UK}

\def\Title#1{\begin{center} {\Large #1 } \end{center}}
\def\Author#1{\begin{center}{ \sc #1} \end{center}}
\def\Address#1{\begin{center}{ \it #1} \end{center}}

\newcommand\pubblock{\rightline{\begin{tabular}{l} \pubnumber\\
         \pubdate  \end{tabular}}}
\newenvironment{Abstract}{\begin{quotation}  }{\end{quotation}}
\newenvironment{Presented}{\begin{quotation} \begin{center} 
             PRESENTED AT\end{center}\bigskip 
      \begin{center}\begin{large}}{\end{large}\end{center} \end{quotation}}

\input econfmacros.tex

\begin{document}
\begin{titlepage}
\pubblock

\vfill
\Title{Commissioning of ELLIE for SNO+}
\vfill
\Author{ Elisabeth Falk$^1$, Jeffrey Lidgard$^2$, Mark Stringer$^1$ and Esther Turner$^2$ (For the SNO+ Collaboration)}
\Address{\sussex}
\Address{\oxford}
\vfill
\begin{Abstract}
SNO+ is a neutrinoless double beta decay and low energy neutrino experiment located in Sudbury, Canada. To improve our understanding of the detector energy resolution and systematics, calibration systems have been developed to continuously monitor the optical properties of the detector, such as: absorption, re-emission, scattering and timing.
\par A part of this in-situ optical calibration system is the Embedded LED/Laser Light Injection Entity (ELLIE). It consists of three subsystems: AMELLIE, SMELLIE, TELLIE. The attenuation module (AMELLIE) is designed to monitor the total optical attenuation, whereas the optical scattering over a wavelength range of 375nm -- 700nm will be characterized by the scattering module (SMELLIE). The timing module (TELLIE) aims to measure the timing characteristics of the photomultiplier tubes.
\par We present the planned commissioning of these three systems, the running of which began early 2017.
\end{Abstract}
\vfill
\begin{Presented}
NuPhys2016, Prospects in Neutrino Physics\\
Barbican Centre, London, UK,  December 12--14, 2016
\end{Presented}
\vfill
\end{titlepage}
\def\thefootnote{\fnsymbol{footnote}}
\setcounter{footnote}{0}

\section{Introduction to SNO+}

SNO+ is a general purpose neutrino detector based 2~km below surface at SNOLAB with the main aim of searching for neutrinoless double beta decay in $^{130}$Te~\cite{gersende}. The detector is a 6~m radius acrylic vessel (AV) filled with detection medium, suspended in a 40~m tall cavern filled with utra-pure water (UPW). The AV is surrounded by $\sim$9300 photomultiplier tubes (PMTs) on a 9~m radius PMT support structure (PSUP)~\cite{whitepaper}. 

There are three phases of detection medium: UPW, pure scintillator and Te loaded scintillator. Currently the AV and cavity are filled with UPW and data taking for the water phase has begun, as has commissioning of the calibration systems~\cite{billy}. 

\section{ELLIE}

The Embedded Laser/LED Light Injection Entity (ELLIE) is part of the in-situ optical calibration system for SNO+. It provides continuous monitoring and aims to improve the understanding of detector energy resolution and systematics. It consists of three subsystems: the Attenuation Module (AMELLIE), the Scattering Module (SMELLIE) and the Timing Module (TELLIE).

\begin{figure}[htb]
\centering
\includegraphics[height=2.5in]{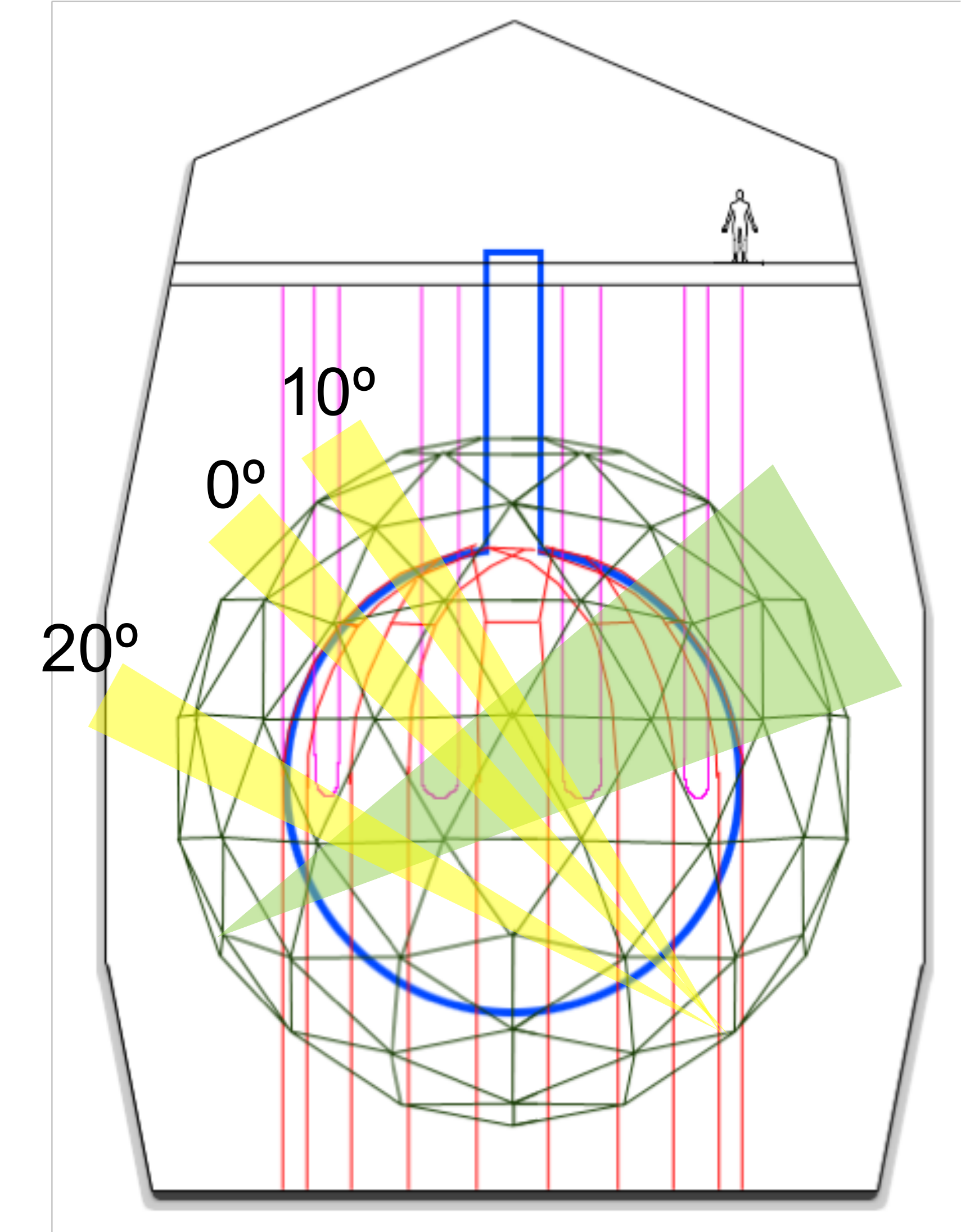}
\caption{A schematic of the SNO+ detector~\cite{krish} with example SMELLIE (yellow) and TELLIE (green) beams overlaid.}
\label{fig:ellie}
\end{figure}

All of these involve optical fibres mounted on the PSUP, connected either to lasers or LEDs  pointed such that the beam passes through the AV, and hence the detection medium, as shown in Figure~\ref{fig:ellie}. These fibres are mounted external to the AV to meet stringent radioactivity requirements.

\section{AMELLIE}

AMELLIE is designed to monitor the attenuation in scintillator, using 8 narrow angle fibres at four injection points, connected to LEDs of two wavelengths. Changes in beam intensity will relate to changing scintillator properties over time. 

In the water phase of the experiment, there are several calibrations to complete, including measurements of angular distributions of beams  as these are an input to simulation and intensity scans in order to calibrate intensity of light to number of photons as seen in the detector. 

\begin{figure}[htb]
\centering
\includegraphics[height=2.5in]{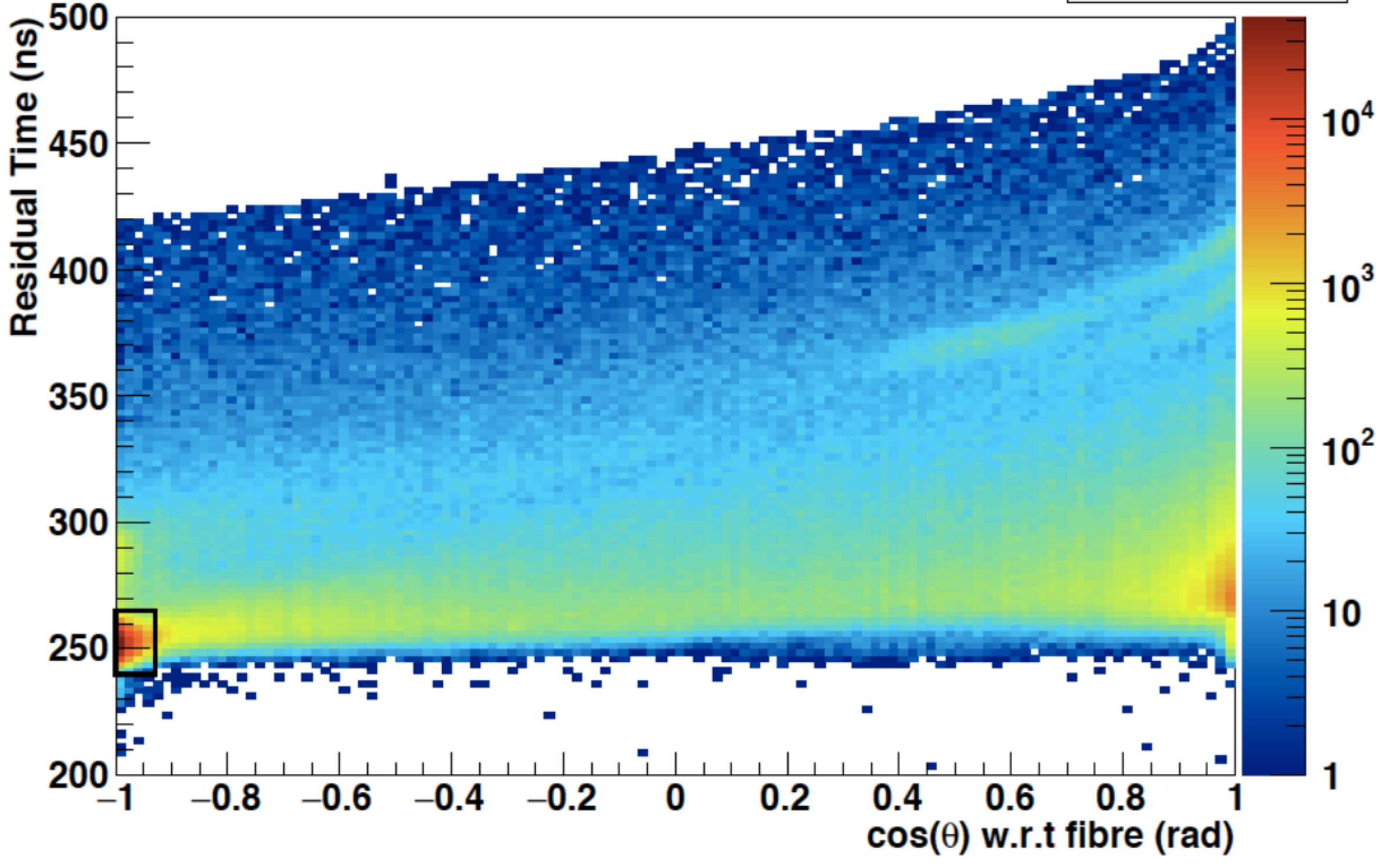}
\caption{Analysis of an simulated AMELLIE beam. Using PMT timing and position information, the direct beam can be identified (red area bottom left). Reflections from other components such as the AV and PMT concentrators can be associated with other features in this plot. }
\label{fig:amellie}
\end{figure}

\section{TELLIE}

TELLIE measures the timing characteristics of the PMTs with 92 light injection points on the PSUP connected to LEDs of 505~nm wavelength. Light is injected with a wide angle, providing full coverage of the PMT array~\cite{tellie}. Therefore, each channel can be used to calibrate timing and gain of the PMTs in its beamspot. By reducing the intensity of light such that each PMT sees a single photoelectron, it is possible to measure the timewalk of the PMTs. 

Additionally, by using light reflected by the AV, it is possible to determine the AV position. From simulation studies, the expected resolution on this measurement is less than 1~cm.

TELLIE has already been fully commissioned. Each channel has been calibrated using a desktop PMT. The three main calibrations of TELLIE are: light output corresponding to each intensity setting; PIN measurement against light output (the PIN reading is a measurement of light intensity made by an internal system); delay between each channel and the trigger out signal.

\begin{figure}[htb]
\centering
\includegraphics[height=2.5in]{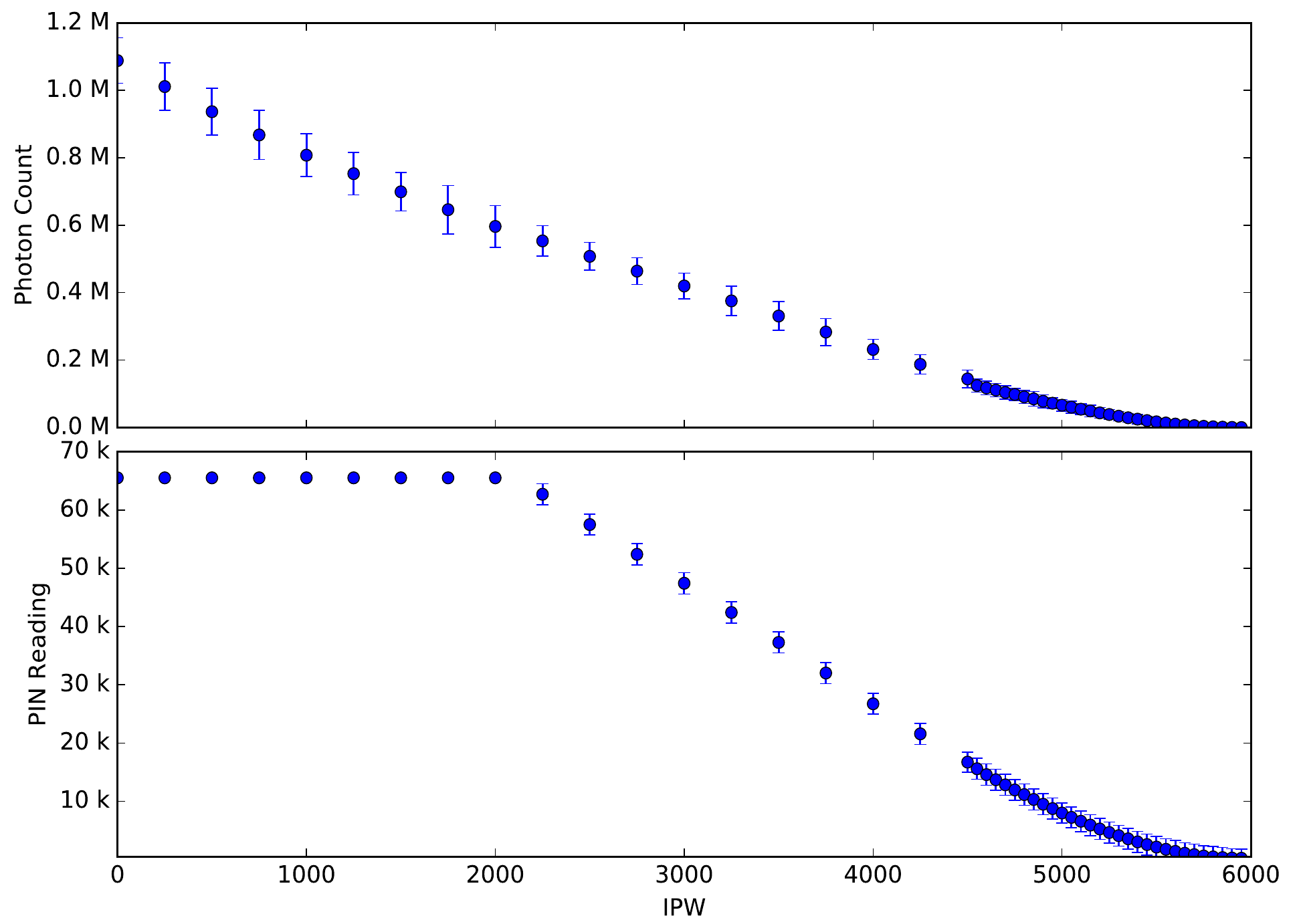}
\caption{Plot of the intensity calibration results for an example channel, with error bars scalled up by a factor of ten for visibility. Parameter IPW sets the intensity of the TELLIE channel. For single photoelectron calibrations, the intensity is set to 1000 photons. The PIN reading saturates at 65536 as it is stored as a 16-bit number and has been calibrated to be sensitive down to low light levels where detector calibration will happen.}
\label{fig:tellie}
\end{figure}

\section{SMELLIE}

SMELLIE is designed to measure and characterise the scattering properties of the detection medium. It consists of 15 collimated fibres at five locations on the PSUP. These are connected to five lasers: four fixed wavelength dye laser heads (375~nm, 405~nm, 445~nm and 495~nm) and one supercontinuum laser (allows wavelengths between 400 -- 700~nm with a bandwidth down to 10~nm). SMELLIE has a monitoring PMT unit (MPU) which is used to measure the intensity of each laser pulse before it enters the detector. This allows for a pulse by pulse correction to the optical intensity of the beam, knowledge of which is relied upon in the analysis technique to determine scattering length.

Water phase is particularly useful for calibration as the scattering properties of water are already well understood. Therefore, there are several calibrations to complete before scintillator is put into the AV. Firstly, the electronics settings of the monitoring system, consisting of the MPU, but also a spectrometer for the supercontinuum laser, need to be optimised. Fibre installation position and angle needs to be confirmed. Then, a map needs to be created of laser intensity settings against wavelength against number of photons entering the detector, for each fibre. This allows the system to be operated with a `multi-hit' correctable region, discussed further in \cite{jack}. Also, detailed measurements of beam profiles, as seen in the detector must be made, as these are an input to the analysis. These calibrations are currently underway. 

\begin{figure}[htb]
\centering
\includegraphics[height=1.5in]{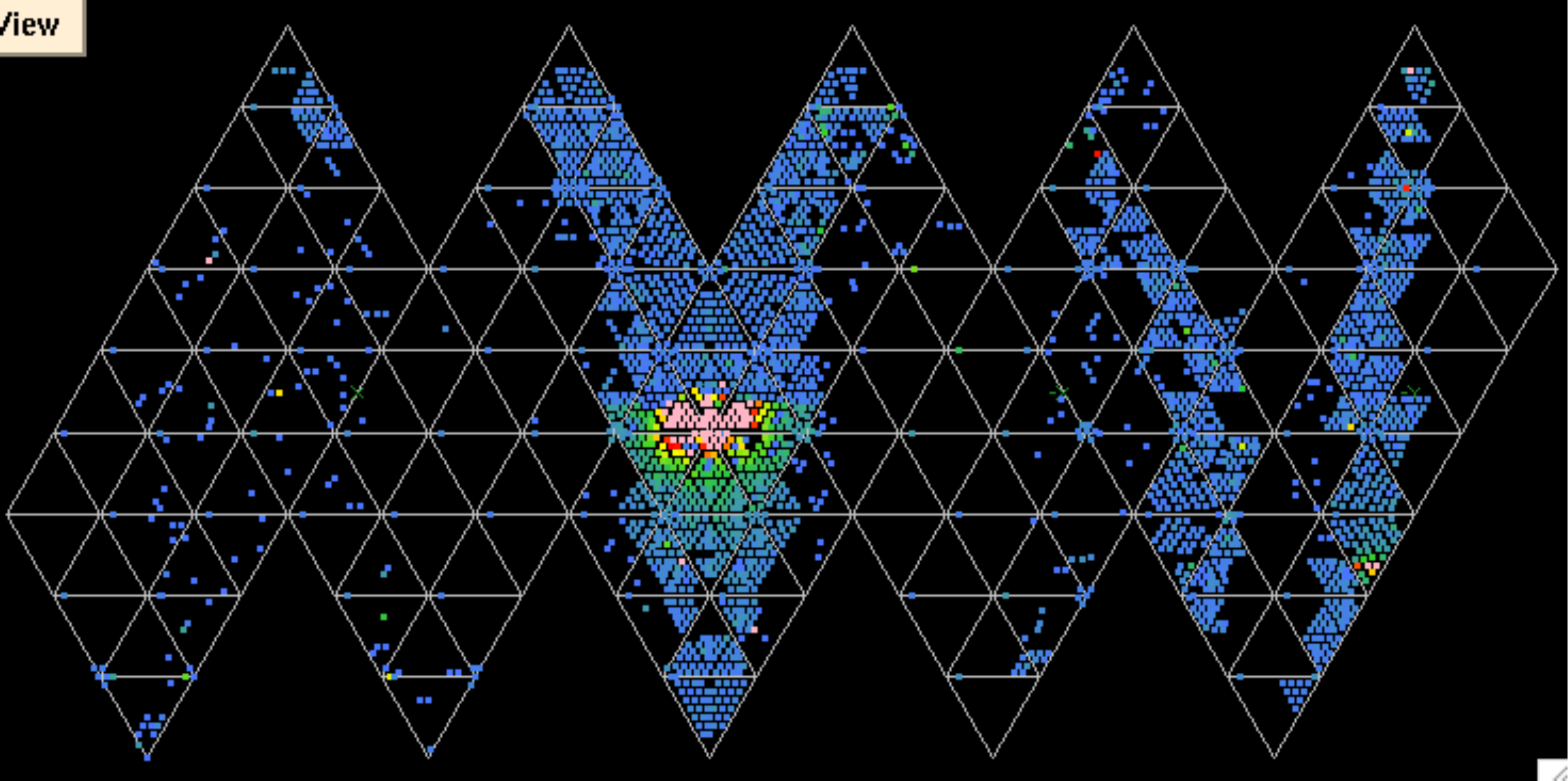}
\caption{Map of PMTs during SMELLIE laser pulses while the detector was partially filled. The direct beamspot is cut in half due to the water level (middle); a reflection from the AV can be seen (bottom right). Colour represents the number of hits, which is proportional to the optical intensity. Only a fraction of PMTs were operating at high voltage in this commissioning run, taken in 2016.}
\label{fig:smellie}
\end{figure}

\end{document}

%% file: econfmacros.tex
\def\beq{\begin{equation}}
\def\eeq#1{\label{#1}\end{equation}}
\def\eeqn{\end{equation}}

\def\beqa{\begin{eqnarray}}
\def\eeqa#1{\label{#1}\end{eqnarray}}
\def\eeqan{\end{eqnarray}}

\let\bar=\overbar

\def\Dslash{\not{\hbox{\kern-4pt $D$}}}
\def\dslash{\not{\hbox{\kern-2pt $\del$}}}

\def\msb{{\bar{\ssstyle M \kern -1pt S}}}

